\documentclass{aa}

\usepackage{graphicx}
\usepackage{txfonts}
\usepackage[mathscr]{eucal}

\begin{document}

\title{Phase-Induced Amplitude Apodization of Telescope Pupils for Extrasolar Terrestrial Planet Imaging}
       
\author{Olivier Guyon}

\offprints{guyon@subaru.naoj.org}

\institute{Subaru Telescope, National Astronomical Observatory of Japan, 650 North A'ohoku Place, Hilo, HI 96720 USA}

\abstract{
In this paper, an alternative to classical pupil apodization techniques (use of an amplitude pupil mask) is proposed. It is shown that an achromatic apodized pupil suitable for imaging of extrasolar planets can be obtained by reflection of an unapodized flat wavefront on two mirrors. By carefully choosing the shape of these two mirrors, it is possible to obtain a contrast better than $10^{9}$ at a distance smaller than $2 \lambda/d$ from the optical axis. Because this technique preserves both the angular resolution and light gathering capabilities of the unapodized pupil, it allows efficient detection of terrestrial extrasolar planets with a 1.5m telescope in the visible.
}
\titlerunning{Phase-Induced Amplitude Apodization of Telescope Pupils for Extrasolar Terrestrial...}
\maketitle

\section{Introduction}
With more than 100 exoplanets now known, direct imaging of extrasolar terrestrial planets (ETPs) is of great scientific importance. Unlike indirect detection techniques, such as radial velocity, astrometry or transit observations, it can allow a spectral decomposition of the light reflected (in the visible) or emitted (in the infrared) by the planet. Such a spectral analysis is essential to study the habitability of ETPs. 

Unfortunately, the large contrast (about $10^{9}$ in visible, and $10^{6}$ in thermal infrared) between ETPs and their parent star, along with the small separation ($0.1''$ for a Earth-Sun system at 10pc) makes direct detection very challenging. Ground-based telescopes of reasonable size (less than 100m diameter) cannot image ETPs in the visible because of the strong wavefront errors due to atmospheric turbulence, and ground-based thermal infrared observations lack the required sensitivity. 

Two solutions are seriously considered to image ETPs from space :
\begin{itemize}
\item{An optical telescope with a high-performance coronagraphic device.}
\item{A thermal infrared ($10 \mu m$) nulling interferometer (\cite{lege96,ange97,guyo02,riau02}).}
\end{itemize}
With an optical telescope, the use of a high-performance coronagraphic device is essential to suppress, or strongly reduce, the amplitude of the ``wings'' of the Point-Spread Function (PSF). Without such a device, the light from the ETPs is ``lost'' in the strong wings of the central star image. 

One such coronagraphic technique, ``classical'' pupil apodization (CPA), aims at modifying the pupil transmission function by placing an amplitude mask in the pupil plane to reduce the amplitude of the PSF wings. This adaptation of the pupil to the problem of high-contrast imaging can allow the PSF wings to be reduced to $10^{-10}$ of the central surface brightness at a distance of a few $\lambda/d$. Many pupil apodization masks have been proposed for this purpose (\cite{jacq64,nise01,aime01,sper01,kuch02}), and optimization techniques to generate these masks have been developed (\cite{gons02}). Unfortunately, this technique reduces the flux sensitivity (a large fraction of the light is absorbed by the apodization mask) and the angular resolution (the transmission at the edges of the apodized pupil is lower than at the center) of the telescope. Moreover, in many CPA designs, only part of the field of view is suitable for planet detection.

In this paper, an alternative to the CPA technique is proposed: phase-induced amplitude apodization (PIAA). Instead of producing the apodized pupil by using an amplitude mask in the pupil plane, the pupil is modified by reflection on 2 mirrors (which can be the primary and secondary mirrors of the telescope) whose phase aberrations are chosen to produce an exit pupil which greatly reduces the intensity of the PSF wings. 

The PIAA technique is first briefly presented in Sect. 2. In Sect. 3, the properties of the images obtained by this technique are studied, and several solutions are presented to use a PIAA telescope as a wide field coronagraphic imager. In Sect. 4, the performance of the PIAA technique for ETPs detection is studied, and the technical feasibility of the technique is discussed in Sect. 5.

\section{Description of the technique}
\subsection{Overview}
\begin{figure}[h]
\centering
\includegraphics[width=9cm]{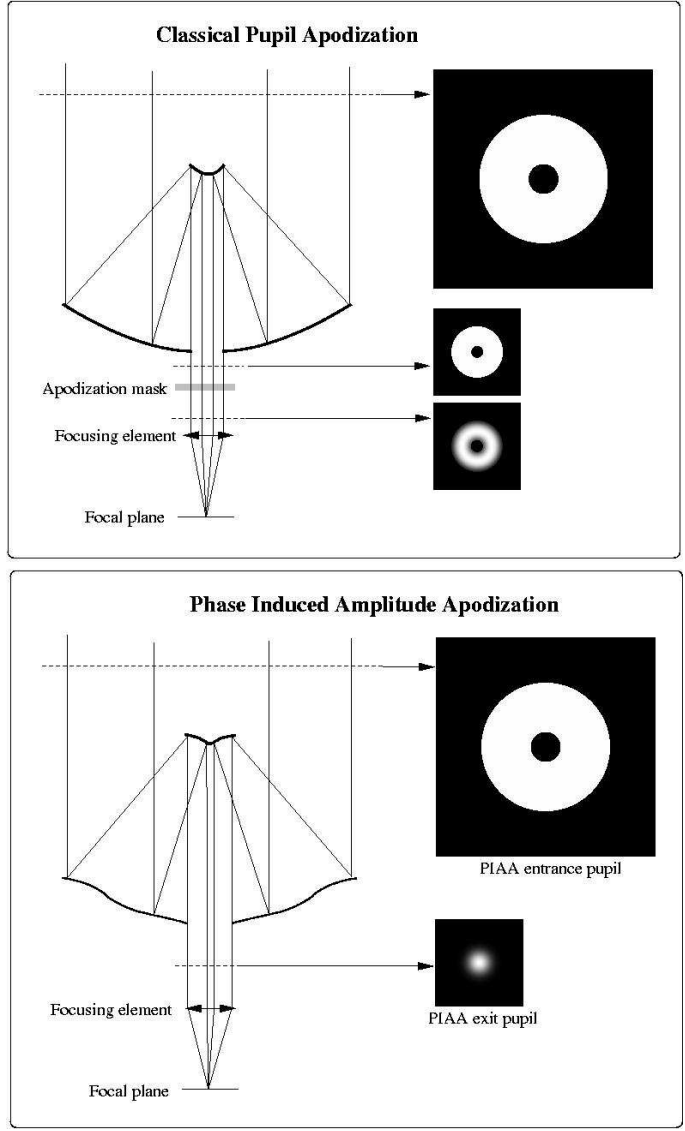}
\caption{Schematic representation of the Classical Pupil Apodization (CPA) technique (top) and the Phase Induced Amplitude Apodization (PIAA) technique (bottom).}
\end{figure}
The general principle of the Phase Induced Amplitude Apodization (PIAA) technique is illustrated in Fig. 1. In a classical telescope, the shape and distribution of light intensity across the pupil is preserved by the reflective optical elements, and only a focus phase term is introduced to form the image on the detector. For clarity, in Fig. 1, this focus phase term is represented by a separate optical element, and consequently, the primary and secondary mirrors' sole function is to reduce the physical size of the beam. To obtain an ``apodized'' pupil with a telescope, a transmissive mask can be placed in a pupil plane (where the wavefronts are flat) as shown in Fig. 1. The PIAA technique uses phase aberrations in two mirrors to produce an apodized pupil. In Fig. 1, the PIAA technique is shown with the primary and secondary mirrors of the telescope used to generate the desired apodized pupil. Two reflections are needed to transform the PIAA entrance pupil into the PIAA exit pupil, which has simultaneously the desired light amplitude distribution and a flat wavefront (aberration-free). 

In the PIAA optical design represented in Fig. 1, the role of the primary mirror is mainly to modify the light distribution across the pupil, while the secondary mirror is used mostly to correct the phase aberrations introduced by the primary mirror. These two separate functions are in reality somewhat shared between the two optical elements. As shown on Fig. 1, the PIAA technique can be used to remove the central obstruction from the pupil.

The PIAA technique is similar to the beam shaping techniques developed to modify laser beam profiles (\cite{shea02}). A similar apodization can also be produced by a secondary mirror designed to correct the spherical aberration of a primary mirror (\cite{gonc02}).

If reflective optics are used (mirrors), the apodized pupil formed by the PIAA technique is perfectly achromatic. This achromaticity is simply due to the achromaticity of the geometric laws of reflection on a mirror. Because the PIAA optics maintain a zero optical pathlength difference across the pupil (see Sect. 2.2.1), the PIAA technique does not introduce phase aberrations, regardless of the wavelength considered. Therefore, the PSF obtained by a PIAA telescope, except for a wavelength proportional scaling factor, is identical at all wavelengths.

\subsection{Computing the mirror shapes}
\subsubsection{Circular-symmetric pupils}
In this section, a geometric method is given to compute the shape of the two mirrors needed to transform any given circular symmetric entrance pupil into the desired circular symmetric exit pupil. Both pupils are free from phase aberrations and are entirely characterized by their radial light distributions, $f_1(r)$ and $f_2(r)$ respectively. The total light flux $F$ is conserved in the PIAA :
\begin{equation}
2 \pi \int_0^{R_1} r\:f_1(r)\:dr = 2 \pi \int_0^{R_2} r\:f_2(r)\:dr = F
\end{equation}
where $R_1$ and $R_2$ are the radius of the primary and secondary mirrors, respectively.

The first step of the geometric construction is to establish a correspondence between the radius in the entrance pupil and the radius in the exit pupil: for a light ray entering the entrance pupil at a radius $r_1$, what is the radius $r_2$ at which this light ray is exiting the exit pupil? This problem is solved by measuring the total flux of the entrance pupil enclosed inside the radius $r_1$, and choosing $r_2$ such that this flux equals the total flux of the exit pupil enclosed inside $r_2$ :
\begin{equation}
2 \pi \int_0^{r_1} r\:f_1(r)\:dr = 2 \pi \int_0^{r_2} r\:f_2(r)\:dr = t \times F
\end{equation}
where $0<t<1$ is the fraction of the light enclosed inside the radii considered.
\begin{figure}[h]
\centering
\includegraphics[width=9cm]{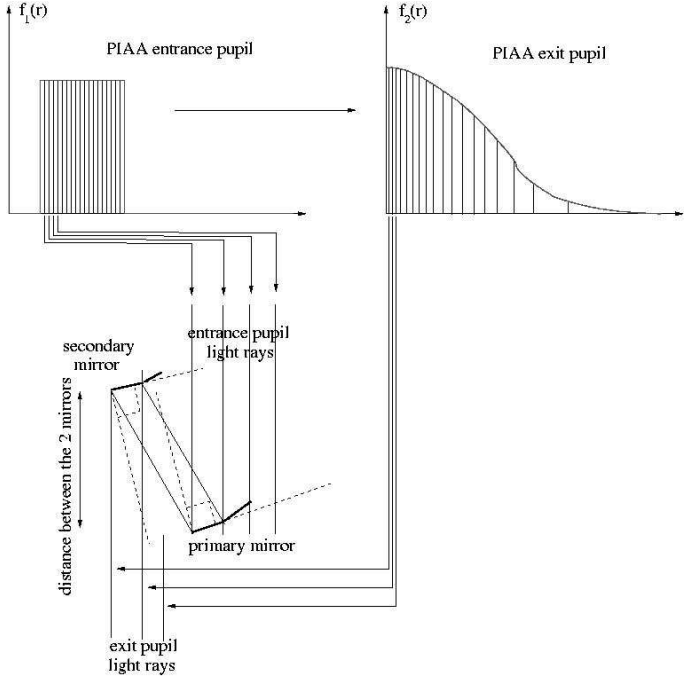}
\caption{A method for geometric construction of the PIAA primary and secondary mirror shapes for circular-symmetric pupils.}
\end{figure}
A large number of radii ${r_1}_i, i=0...N-1$, of increasing values, is chosen to sample the entrance pupil, and the corresponding radii ${r_2}_i$ are computed for the exit pupil. The entrance pupil and exit pupil light rays can then be drawn (see Fig. 2). The path of the the first light ray can be drawn because the position of the points at which this ray hits the primary and secondary mirrors are known from the distance between the 2 mirrors. For each mirror, the angle of incidence of this first light ray with the surface of the mirror is measured (it is half of the angle between the incoming and reflected light rays). This angle of incidence gives the slope of the mirror surface at the reflection point, and a small mirror segment can then be constructed up to the point where the next light ray is encountered. This process is repeated until the last light ray, and leads to an approximation of the mirror shapes made of $N$ straight segments. By increasing the number of sampling rays $N$, the obtained mirror shapes $M_1(r)$ (primary mirror) and $M_2(r)$ (secondary mirror), by construction, satisfy the following conditions :
\begin{itemize}
\item{There is no phase discontinuity across the outgoing wavefront, and both the incoming (entrance pupil) and outgoing (exit pupil) light rays are parallel to each other. Therefore, the incoming and outgoing wavefronts are flat (a direct consequence of Huygens-Fresnel principle).}
\item{The reflections of the incoming light rays on $M_1$ and then $M_2$ produce the desired exit pupil light distribution $f_2(r)$.}
\end{itemize}
Software was written to compute $M_1(r)$ and $M_2(r)$ from $f_1(r)$ and $f_2(r)$, and, to check the validity of the obtained solution, the optical pathlength difference $OPD_i$ between rays $i$ and $0$ was computed for $i=1...N-1$. It was found that the residual OPD errors decrease toward 0 when $N$ increases:
\begin{equation}
\lim_{N \longrightarrow \infty} \left(\sum_{i=1}^{N-1} (OPD_i)^2\right) = 0.
\end{equation}
If $f_2(r) = \alpha^2 \times f_1(\alpha \times r)$ (the exit pupil is identical to the entrance pupil, scaled by a factor $\alpha$), then the software correctly produces a primary mirror $M_1$ which is a parabola.

This geometric construction is equivalent to solving the following differential equation:
\begin{equation}
\frac{d\:M_1(r_1(t))}{d\:r_1(t)} = \frac{d\:M_2(r_2(t))}{d\:r_2(t)} = \frac{\sqrt{A(t)^2+B(t)^2}-B(t)}{A(t)}
\end{equation} 
where
\begin{equation}
A(t) = r_2(t)-r_1(t),
\end{equation}
\begin{equation}
B(t) = M_2(r_2(t))-M_1(r_1(t)),
\end{equation}
and $r_1(t)$ and $r_2(t)$ are respectively the radii in the entrance and exit pupils inside which the fraction of the total flux is $t$ (Eq. (2)). The initial conditions are the size of the central obstruction ($r_1(0)$), if any, and the distance between the primary and secondary mirror ($M_2(0)$, assuming that $M_1(0)=0$). The first equality in Eq. (4) is ensuring that the PIAA exit wavefront is flat: for a given light ray (parameter $t$), the angles of the two reflections are opposite and produce an output light ray parallel to the incoming light ray.

\subsubsection{Off-axis telescopes with circular symmetric entrance and exit pupils.}
High dynamical range imaging, whether with the CPA or PIAA techniques, is easier without the central obstruction and the ``shadow'' of the structure that supports the secondary mirror (spider). Although the PIAA technique can ``erase'' the central obstruction (as illustrated in Fig. 1), a large central obstruction produces a greater mismatch between $f_1(r)$ and $f_2(r)$ which, as will be shown later, reduces the imaging field of view. Moreover, removing the spider with the PIAA technique can be difficult.

Just as in the case of ``classical'' telescopes, the mirror shapes computed by the technique presented above can be modified to be used in an off-axis telescope by simply adding a constant slope across the mirror. In the case of a classical telescope, this operation transforms a parabola into an off-axis parabola ($x^2+\alpha \times x = (x+\alpha/2)^2-\alpha^2$), and it can be applied on both the primary and secondary mirrors. It is easy to check that this operation only ``steers'' the whole beam, but does not introduce phase aberrations or modifies the light amplitude distribution in planes perpendicular to the central light ray.

\subsection{One example}
In this section, the PIAA technique is illustrated by an example. The desired radial light intensity $f_2(r)$ to be obtained by the PIAA technique is shown in Fig. 3, along with the radial profile of the PSF it produces. The function $f_2$ was computed through the iterative algorithm presented in Sect. 4.1, and is optimized to reduce the intensity of the PSF at distances greater than $\lambda/d$. The strict constraints put on the PSF produced a function $f_2$ which is null on four intervals separated by intervals in which $f_2$ is strictly positive. Very good apodization functions which have no such null intervals exist, and are also suitable for ETPs detections. 
\begin{figure}[h]
\centering
\includegraphics[width=9cm]{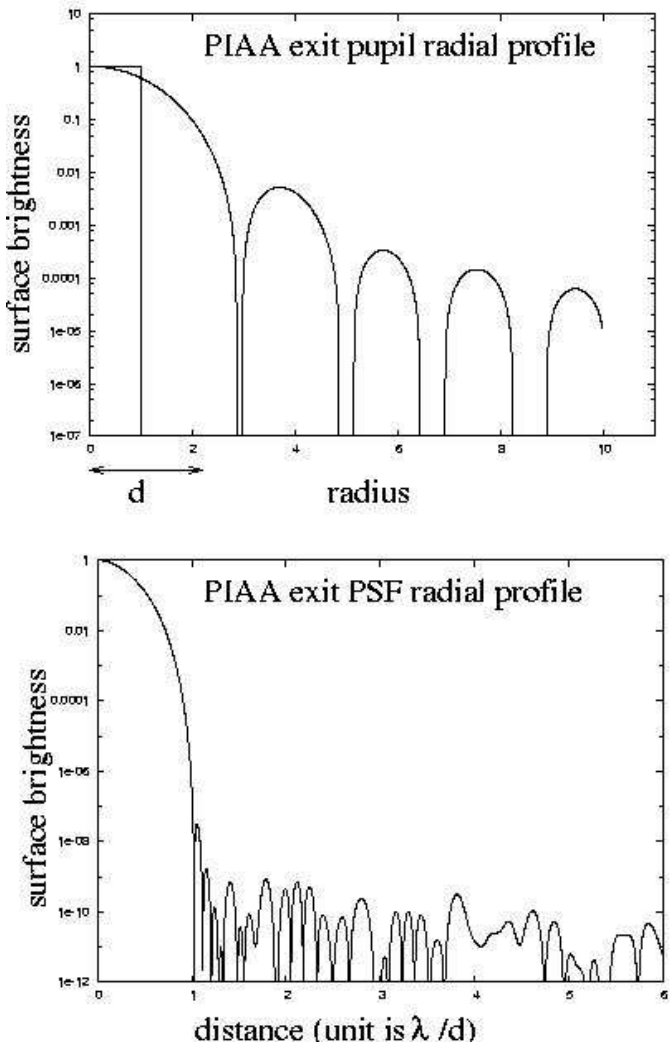}
\caption{Radial profile $f_2(r)$ of the PIAA exit pupil (top) and radial profile of the corresponding PSF (bottom) for the example considered. The radial profile of the unapodized pupil of equal total flux is shown as a dashed line, and its diameter $d$ is used to measure the distances in the PSF (unit is $\lambda/d$).}
\end{figure}

\begin{figure}[h]
\centering
\includegraphics[width=9cm]{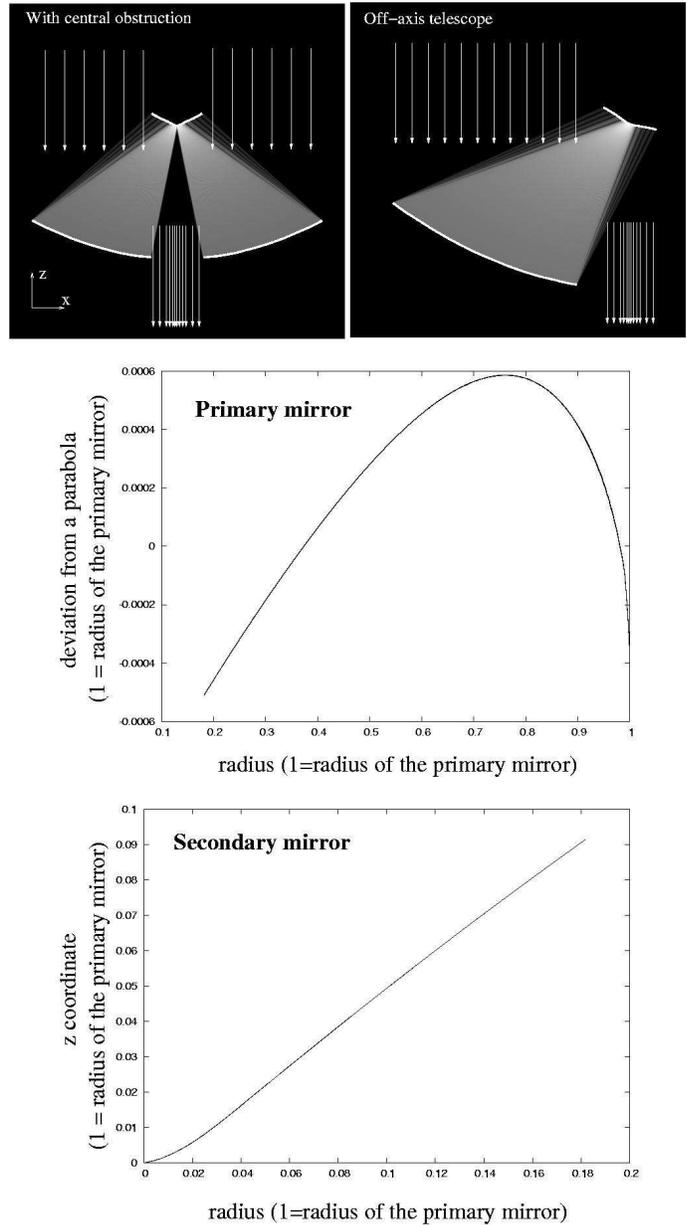}
\caption{Optical designs for the PIAA example studied: telescope with central obstruction (upper left) and off-axis telescope (upper right). Both images show a cut of the mirror shapes and the density of light rays between the two reflections. A few light rays are represented entering the telescopes and the corresponding light rays exiting the telescope are shown. The shape of the optics is shown for the telescope with central obstruction: deviation of the primary mirror to a parabolic fit (center) and shape of the secondary mirror (bottom).}
\end{figure}

Figure 4 shows two possible optical configurations: on-axis telescope or off-axis telescope. In both cases the distance between the 2 mirrors is less than the diameter of the primary, and the primary is therefore close to a fast parabola ($F/D<1$), as shown in Fig. 4, center: for a 2m diameter on-axis primary mirror, the difference between the primary mirror and a perfect parabola is less than $1\:mm$. This deviation from a perfect parabola is mostly seen in the outer edge of the primary, where the surface if ``bent'' to reflect a small amount of the light into the outer part of the secondary (the ``wings'' of the $f_2$ function). The shape of the secondary mirror (Fig. 4, bottom) is also peculiar, and its central part is approximated by a cone (the slope of the mirror does not tend to 0 at small distance from the optical axis) in the case of a telescope with central obstruction. 

\section{Field of view and possible use as a coronagraph}
\subsection{Field of view of the technique}
\begin{figure}[h]
\centering
\includegraphics[width=9cm]{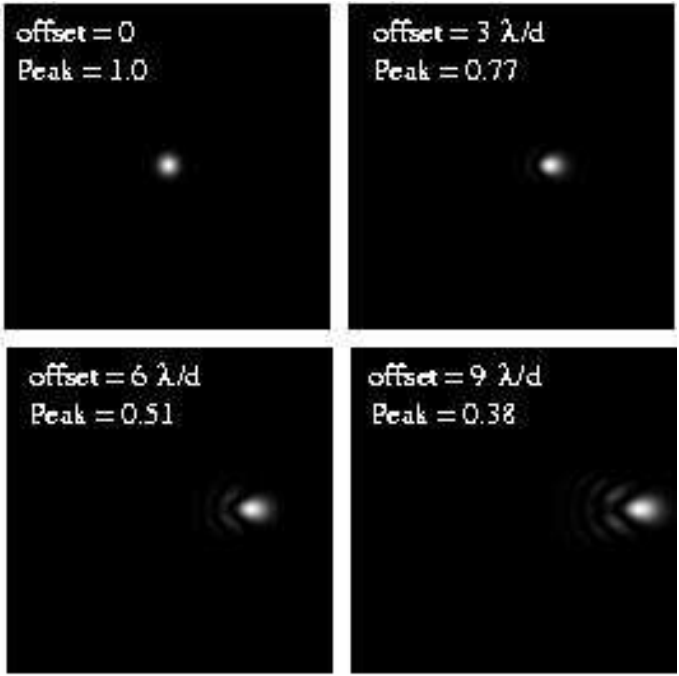}
\caption{Degradation of the PSF quality with distance to the optical axis for the example studied in Sect. 2.3. The brightness scale is linear and identical for all images.}
\end{figure}
In a classical telescope, an off-axis source produces the same wavefront as an on-axis source modulo a phase slope. This slope is given by the position of the source relative to the optical axis. As a result, the image of an off-axis source is obtained by translation of the image of an on-axis source.

The PIAA technique redistributes the complex amplitude of the light in the entrance pupil to form the PIAA exit pupil. The incoming wavefront of an on-axis point source is flat and its phase is constant across the pupil, and because the PIAA does not introduce phase aberration, this property is still true in the PIAA exit pupil. However, the geometric redistribution of light between the entrance pupil and the exit pupil of the PIAA does not preserve the constant phase slope of a wavefront from an off-axis source. The PSF is therefore not translation-invariant as shown in Fig. 5: the PSF becomes less concentrated for off-axis sources.

This effect does not affect the performance of a PIAA imager for detection of ETPs within a few times $\lambda/d$ because the loss of PSF quality is moderate to small in this central region, and the image of the star (defined by the PSF of an on-axis source) remains very contrasted (contrast better than $10^{9}$ at $2\times \lambda/d$ in the example considered). However, wide field imaging performance of the PIAA imager is seriously affected by this effect. In the particular example studied, the point source detection limit for planets at large angular separation would be affected, especially in the presence of a strong background (exozodiacal light for example).

\subsection{Extending the field of view of the PIAA imager}
The first solution to the field of view problem is to carefully choose the light distribution of the PIAA exit pupil to maintain a bright, sharp and contrasted PSF core even at large angular separations from the optical axis. This can be done by imposing additional constraints in the iterative algorithm used to compute $f_2(r)$ (see section Sect. 4.1), in order to minimize the mismatch between $f_1(r)$ and $f_2(r)$. For example, if the entrance pupil of the PIAA is an unobstructed disk, as much of the light in the PIAA exit pupil as possible should be within a large ``flat'' central part of $f_2(r)$: this ``flat'' part will produce a bright and sharp diffraction core while the wings of $f_2$ will be used to cancel the diffraction and yield a very contrasted on-axis PSF.

A hybrid CPA/PIAA optical configuration is also possible. For example, the central part of the $f_2$ function can be obtained by classical apodization while the wings of the $f_2$ function are obtained by the PIAA technique. This hybrid configuration would extend the field of view at the expense of a loss of light.

\subsection{Use as a coronagraph}
The PIAA entrance pupil (telescope pupil) offers a wide field of view but a low contrast PSF, while the PIAA exit pupil offers a good contrast PSF but a small field of view. It is possible to combine the advantages of these two pupils by taking advantage of the PSF obtained by the PIAA exit pupil (apodized) to mask the central source and then restore the original pupil (telescope pupil) before forming the final image. Figure 6 illustrates this optical configuration.

\begin{figure}[h]
\centering
\includegraphics[width=9cm]{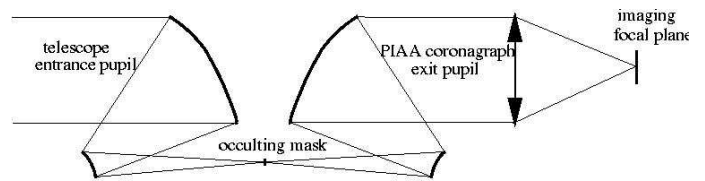}
\caption{Schematic representation of the PIAA coronagraph.}
\end{figure}

If the occulting mask is removed, the effect of the 4 mirrors cancel each other and the PIAA coronagraph exit pupil is identical to the telescope entrance pupil. The PSF is therefore translation-invariant and of good quality across a wide field of view. An off-axis source is not affected by the occulting mask because its image misses the small mask in the first focal plane, but the central source's light is blocked by the occulting mask, whose size is chosen to match the size of the on-axis PSF.

The optical quality requirements are much more strict for the two mirrors before the mask than after the mask: the PSF wings must be kept very faint at the first focal plane. Therefore, the addition of two extra optical elements to build a PIAA coronagraph does not pose serious difficulties, and the wavefront accuracy for these two mirrors can be as low as $\lambda/10$ without significant loss of performance.

This technique is very similar to the coronagraphic technique used in the interferometer concept studied by Guyon \& Roddier (2002), where pupil densification (\cite{labe96}) is used to first adapt the interferometer's sparse pupil to a coronagraph, and pupil redilution is then used to restore the entrance's pupil shape. This later step is essential to restore the wide field of view that is lost in the pupil densification process.

\section{Performance}
\subsection{Choice of the apodization function}
The performance of the PIAA imager/coronagraph is solely a function of $f_2$, the light distribution in the PIAA exit pupil. The constraints on the choice of $f_2$ are first discussed :
\begin{itemize}
\item{{\bf (1) Positivity of $f_2$.} In the technique presented in this paper, there is no convenient way to introduce an achromatic $\pi$ phase shift in parts of the pupil, and $f_2$ is therefore real and positive.}
\item{{\bf (2) Contrast of the on-axis PSF.} The main requirement for the detection of ETPs is to obtain a PSF which offers high contrast (typically $10^9$ to $10^{10}$) at small angular distances (one to two times $\lambda/d$).}
\item{{\bf (3) Finite size of the secondary mirror.} The finite size of the secondary mirror and the fact that the ``core'' of the PIAA exit pupil cannot be made arbitrarily small (because of Fresnel diffraction) imposes a limit on the extent of $f_2$.}
\item{{\bf (4) Field of view considerations.} This constraint has been shown in Sect. 3, and requires a $f_2$ function which is somewhat flat in its central part (small values of $r$). The PIAA coronagraph does not have this constraint.}
\end{itemize}

\begin{figure}[h]
\centering
\includegraphics[width=9cm]{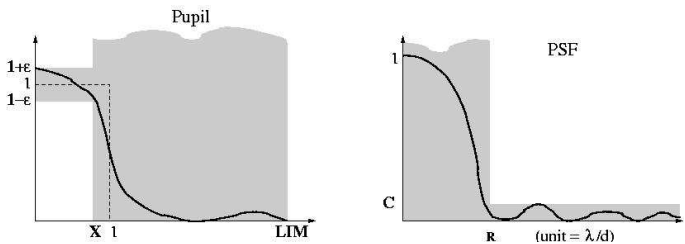}
\caption{Constraints imposed on the function $f_2$ and on the on-axis PSF. The grey area represents the permitted values for both $f_2(r)$ and the PSF radial profile and the thick curves represent examples of permitted functions. In the pupil radial profile, the dashed line represents the unapodized unobstructed pupil of diameter d, and of total flux equal to the total flux of the PIAA exit pupil.}
\end{figure}

Figure 7 shows how these constraints were applied with five parameters:
\begin{itemize}
\item{$\epsilon$ : maximum deviation of $f_2$ from the flat reference pupil.}
\item{$X$ : radius of the PIAA exit pupil inside which the maximum deviation of $f_2$ from the flat reference pupil is enforced.}
\item{$LIM$ : maximum size of the PIAA exit pupil.}
\item{$C$ : contrast to be achieved by the PSF.}
\item{$R$ : radial distance outside of which the contrast to be achieved by the PSF is enforced.}
\end{itemize}

These constraints are used in a Gerchberg-Saxton (\cite{ger72}) type iterative algorithm to find a function $f_2$ which satisfies all constraints. At each iteration, the Fourier transform of the pupil (the focal plane light complex amplitude) is first computed, and the PSF constraints (2) are used to modify the focal plane light complex amplitude. Another Fourier transform is performed to compute the pupil light complex amplitude distribution, which, in turn, is modified by the constraints (1), (3) and (4). After several iterations, convergence of the pupil light complex amplitude distribution toward a suitable function $f_2$ usually occurs. If no convergence occurs, the constraints parameters are relaxed (for example, larger size of the secondary mirror, or lower required contrast of the PSF) until convergence occurs. Because this study is limited to radially symmetric pupils, all computations are actually done in one dimension.

The example shown in section Sect. 2.3 was obtained by this algorithm with a set of parameters which yields a PSF suitable for ETPs detection in the imager configuration ($\epsilon = 0.25$, $X = 0.8$, $LIM = 10$, $C = 10^{-9}$, $R = 1.0$). Although the field of view of this particular example is well suited for ETPs detection with a small telescope ($d \approx 2m$), wide field imaging would require either a different choice of values for $\epsilon$ and $X$, or a coronagraphic configuration.

\subsection{Performances and comparison with classical apodized pupils}
Classical apodization techniques (CPA) have several disadvantages compared to the PIAA technique:
\begin{itemize}
\item{{\bf Loss of flux}. The transmission of classical apodization masks is usually between 0.1 and 0.5, while no light is absorbed with the PIAA.}
\item{{\bf Loss of angular resolution.} In CPAs, the apodization mask usually has a lower transmission at the edges of the pupil, resulting in a loss of angular resolution by a factor of approximately 2. This loss of resolution requires the use of a larger telescope and also results in lower sensitivity, since more background light (from scattering of the central source's light or emission from the zodiacal/exozodiacal clouds) is mixed with the companion's PSF.}
\item{{\bf Limited useful field of view.} Most apodization masks produce a PSF in which only a fraction of the field is usable for ETPs detection.}
\end{itemize}
When choosing an apodization mask, CPA techniques must find an acceptable tradeoff between these three different losses, and minimization of one of these losses usually comes at the expense of at least another loss.

To estimate the performance of the PIAA and CPA techniques for direct planet imaging, the detection signal to noise ratio (SNR) is computed for a representative case, with the diameter of the telescope as the main variable. Four configurations were tested under the same conditions :
\begin{itemize}
\item{{\bf PIAA configuration.} The example studied in Sect. 2.3 was adopted.}
\item{{\bf Jacquinot configuration} (\cite{jacq64}). The pupil transmission is equal to 1 if and only if:
\begin{equation}
|y| < R \times \left( e^{-\left(\frac{\alpha x}{R}\right)^2}-e^{-\alpha^2}\right)
\end{equation}
where $R$ is the radius of the pupil, $x$ and $y$ are the 2D coordinates in the pupil plane, and $\alpha = 2$. The pupil transmission is equal to 0 everywhere else.
}
\item{{\bf Spergel configuration} (\cite{sper01}).  
The pupil transmission is equal to 1 if and only if:
\begin{equation}
y_0(x) < |y| < y_0(x)+y_w(x)
\end{equation}
with
\begin{equation}
y_w(x) = R \times \left( e^{-\left(\frac{\alpha x}{R}\right)^2}-e^{-\alpha^2}\right)
\end{equation}
and
\begin{equation}
\int_{y_0(x)}^{y_0(x)+y_w(x)} y^2 dy = \beta y_w(x)
\end{equation}
where $\alpha=2$, $x$ and $y$ are the 2D coordinates in the pupil plane, and $\beta=0.4 \times R^2$. The pupil transmission is equal to 0 everywhere else.
}
\item{Apodized Square Aperture (ASA) configuration (\cite{nise01}). The pupil transmission is 0 everywhere except in a square of radius $R$, inside which it is equal to:
\begin{equation}
P(x,y) = \left(1-\left(\frac{x}{R}\right)^2\right)^\nu \times \left(1-\left(\frac{y}{R}\right)^2\right)^\nu
\end{equation}
where $\nu=5$.
}
\end{itemize}
The 3 CPA pupils adopted for the comparison are shown in Fig. 8, along with the corresponding PSFs.
\begin{figure}[h]
\centering
\includegraphics[width=9cm]{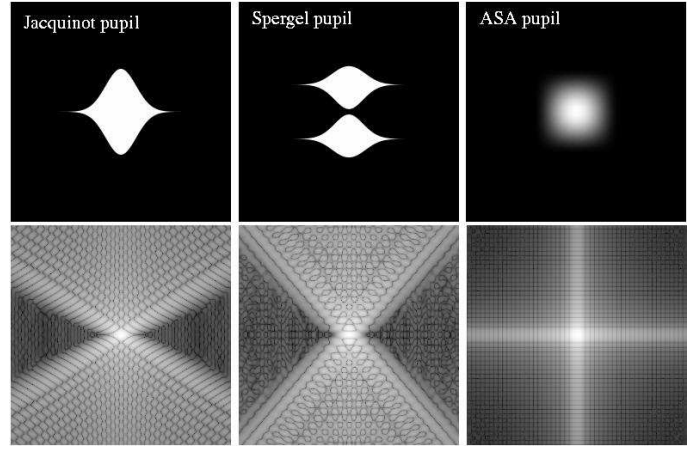}
\caption{CPA pupils adopted for the comparison between PIAA and CPA technique performance. The pupils are shown on the upper row and the corresponding PSFs are shown in the lower row (logarithmic scaling).}
\end{figure}

To compute the SNR of a companion detection, only photon noise is considered, and the noiseless PSF is supposed to be perfectly known. For a given image of the star/planet system, the best detection SNR is obtained by computing the optimally weighted sum of the PSF-subtracted image pixel values. The weight of each pixel in the sum is proportional to the square of the SNR on that pixel and the resulting detection SNR is
\begin{equation}
SNR = \sqrt{\int_{x,y} \left(\frac{I_c(x,y)}{\sqrt{I_c(x,y)+I_s(x,y)}}\right)^2 dx dy}
\end{equation}
where $I_c(x,y)$ is the image of the companion and $I_s(x,y)$ is the image of the central star. In the case of the PIAA configuration, the PSF is not translation-invariant, and has to be computed both for the central star and the planet. For a given optical configuration, this SNR is {\bf a} function of the contrast between the star and the companion, their angular separation, the luminosity of the star and the diameter of the telescope. For PSFs which are not rotational-symmetric, the SNR is also a function of the position angle (PA), in which case the detection SNR can be written
\begin{equation}
SNR = \sqrt{\int_{PA=0}^{2 \pi } \left(SNR(PA)\right)^2 \:dPA}
\end{equation} 
where $SNR(PA)$ is computed by Eq. (10). For the Jacquinot and Spergel pupils, where the PSF contrast is very good only for a range of PA values, this equation is equivalent to rejecting the ``unusable'' regions of the PSF (too low SNR).

The four optical configurations are tested on a $m_v=5$ star with a planet $10^9$ times fainter at a separation of 0.1''. The central wavelength is $0.5 \mu m$ and the spectral bandwidth is $0.2 \mu m$. No zodiacal or exozodiacal light was included in this simple simulation, and the detectors are assumed to be perfect (no readout noise).

The results of the performance analysis is shown in Fig 9. 
\begin{figure}[h]
\centering
\includegraphics[width=9cm]{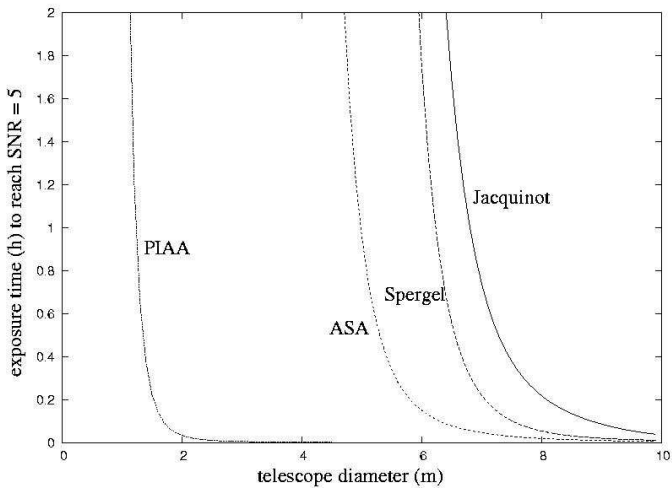}
\caption{Telescope diameter needed to detect (SNR=5) a companion 0.1'' from a $m_v=5$ star with a flux ratio of $10^9$. The central wavelength is $0.5 \mu m$ and the bandwidth is $0.2 \mu m$.}
\end{figure}
The 3 CPA configurations require telescopes of diameters between 5 and 7 meters to detect (SNR=5) the ETP in one hour exposure time, while the PIAA configuration reaches the same performance with a 1.5m diameter telescope. This result is explained by careful analysis of the table 1.

\begin{table}[h]
\begin{tabular}{|c|c|c|c|}
\hline
 & Transmission & IWD & Discovery Space \\
\hline
\hline
PIAA & 1.00 & $\approx 1.5 \lambda/d$ & 1.00 \\
\hline
Jacquinot & 0.285 & $\approx 5 \lambda/d$ & $\approx 0.25$\\
\hline
Spergel & 0.273 & $\approx 4 \lambda/d$ & $\approx 0.5$ \\
\hline
ASA & 0.107 & $\approx 4 \lambda/d$ & $\approx 0.5$\\
 & & & (increases \\
 & & & with distance)\\
\hline
\end{tabular}
\caption{Main properties of the PIAA configuration and 3 CPA configurations. IWD stands for Inner Working Distance, and is the smallest angular distance at which a $10^9$ contrast is reached in the PSF. The Discovery space is the fraction of the field of view usable for faint companion detection.}
\end{table}

The Inner Working Distance (IWD), which is the angular separation at which the contrast of the PSF reaches $10^9$, is about 3 times larger for the CPAs than for the PIAA configuration. This means that, in the visible, the minimum telescope diameter required to detect ETPs in reasonable exposure times is about $4$ to $5$ meters for CPA configurations while it is only $1.5$ meter for the PIAA configuration. In addition, the low transmission and small discovery space of the CPA configuration yield a comparatively lower efficiency compared to the PIAA configuration. The product of the transmission by the discovery space is a good approximation of this efficiency: for the CPA configurations, this efficiency is about 0.1 (responsible for an increase of the exposure time by a factor of 10).

The fact that ETPs can be detected with a 1.5m telescope is not surprising when the number of incoming photons is considered. In the typical example studied above, the ETP photon flux is about $0.2 \: photons/s/m^{2}$, and about 1200 photons can be collected by a 1.5m telescope in 1 hour.

\section{Technical challenges and feasability}

\subsection{Diffraction}
Fresnel propagation of a collimated beam over large distances modifies its light intensity distribution. This is especially true if the beam has sharp edges, such as the boundaries created by the edge of the primary mirror and the support structure of the secondary mirror (central obstruction and spider) in an on-axis telescope. In the PIAA technique, the beam is strongly apodized (no sharp edges) from the second PIAA mirror (secondary mirror in Fig. 1) to the final focal plane. The effect of diffraction on the performance of the PIAA telescopes is therefore not as serious as it is for non-apodized coronagraphic telescopes, and can be kept acceptable for ETPs detection by :
\begin{itemize}
\item{{\bf Adopting an off-axis telescope configuration.} The PIAA entrance pupil is then a full disk, and the effect of diffraction is limited to the edges of this disk.}
\item{{\bf Keeping beam lengths small.} This is especially true if a collimated beam is transported to small-size PIAA optics. If the unobstructed collimated beam is propagated on a long distance, its edges will be distorted (amplitude and phase) by Fresnel propagation. Both the F/D ratio of the telescope and the ratio between the propagation length and the diameter of the beam should be kept reasonably low.}
\item{{\bf Controlling the diffraction.} If diffraction from the edges of the pupil is a problem, it can usually be suppressed by slightly diaphragming (masking the edges of the pupil) the beam immediately before the PIAA optics.}
\end{itemize}

\subsection{Quality of the optics}
The requirements on the quality of the optical elements in a telescope for ETPs detection are similar for the PIAA and CPA techniques. 

The phase of the exit wavefront (the apodized wavefront in case of the PIAA and CPA techniques) needs to be flat to within approximately $\lambda/3000$ for efficient ETPs detection (\cite{kuch02}). Larger wavefront errors can be tolerated at the expense of increased exposure times, if the wavefront is still stable enough to allow an accurate calibration of the PSF. Scattering by the optics also needs to be very low, especially in the mirrors that are the closest to the PIAA exit focal plane, and good spatial uniformity of the reflectivity of the mirrors is essential.

Since the polishing of full-size (1.5 meter diameter primary) PIAA optical elements is currently very challenging, an optical configuration with a ``classical'' off-axis afocal telescope feeding small PIAA optics is probably preferable. In this optical configuration, the required optical quality of the telescope optics (primary and secondary mirrors) is identical to the required optical quality for CPA telescopes or more ``classical'' coronagraphs aimed at detecting ETPs. However, the smaller size (by a factor of 3) of the telescope allowed by the use of the PIAA technique makes the polishing of these optical elements significantly easier. The even smaller size of the ``apodizing'' mirrors, which could be only a few centimeters in diameter, makes it possible to efficiently use high accuracy polishing techniques, such as ion beam figuring. The polishing of the PIAA optics offers similar difficulties as the polishing of ``wild aspherics'' optical element in imaging systems.

\section{Discussion}
The PIAA technique makes it possible to image ETPs with a telescope smaller than 2m in less than 1h exposure time, which makes it a very attractive alternative to classical pupil apodization (CPA) techniques. A direct comparison between the two techniques shows that a reduction by a factor of three in telescope diameter can be gained by using the PIAA configuration. 

The PIAA technique can also be used to adapt the telescope's pupil to a coronagraph. For example, it can produce a pupil suitable for the phase mask coronagraph (\cite{rodd97}), which can only reach full coronagraphic extinction with a suitably apodized pupil (\cite{guyo00,aime02,guyo02,soum03}). This pupil adaptation to the coronagraph can also be performed on the densified pupil of an interferometer.
Another interesting application is to increase the coupling efficiency between a telescope and a single mode fiber in interferometers. By producing a suitable unobstructed apodized exit pupil, the PIAA technique can increase the coupling efficiency by 22\% for an unobstructed telescope pupil, and by 93\% for the CFHT pupil (0.44 central obstruction).



The PIAA technique could also be used on a smaller size telescope (about 0.5m to 1m diameter), or at a longer wavelength, to detect Jupiter-like planets. The technique could also replace Lyot coronagraphs on ground-based telescopes with adaptive optics for scientific projects requiring lower contrast ratios. The PIAA technique could be implemented with two small high-order deformable mirrors on ``classical'' telescopes, and these two mirrors could simultaneously correct the wavefront aberrations of the telescope's optics and reduce the intensity of the PSF diffraction wings.

\begin{acknowledgements}
The author is thankful to Katherine Roth and Thomas Kane for suggestions to improve the manuscript, and to Lyu Abe and Koji Murakawa for discussions about this idea. The author is also thankful to the referee for many useful comments.
\end{acknowledgements}


\begin{thebibliography}{}
\bibitem[Aime et al., 2001]{aime01} Aime, C., Soummer, R., \& Ferrari, A.  2001, A\&A, 379, 697.
\bibitem[Aime et al., 2002]{aime02} Aime, C., Soummer, R., \& Ferrari, A.  2002, A\&A, 389, 334.
\bibitem[Angel \& Woolf, 1997]{ange97} Angel, J.R.P., \& Woolf, N.J.  1997, ApJ, 475, 373.
\bibitem[Gerchberg \&  Saxton, 1972]{ger72} Gerchberg, R.W., \& Saxton, W.O.  1972, Optik, 35, 237.
\bibitem[Goncharov \& Puryayev, 2002]{gonc02} Goncharov, A.V.,\&  Puryayev, D.T.  2002, Opt. Eng., 41, 3111. 
\bibitem[Gonsalves \& Nisenson, 2002]{gons02} Gonsalves, R., \& Nisenson, P.  2002, preprint, (astro-ph/0210166)
\bibitem[Guyon \& Roddier, 2000]{guyo00} Guyon, O., \& Roddier, F.  2002, Proc. of the conf. ``Darwin and Astronomy - The infrared Space Interferometer'', Stockholm, Sweeden, 17-19 November 1999, 41.
\bibitem[Guyon \& Roddier, 2002]{guyo02} Guyon, O., \& Roddier, F.  2002, A\&A, 391, 379.
\bibitem[Jacquinot \& Roizen-Dossier, 1964]{jacq64} Jacquinot, P., \& Roizen-Dossier, B.  1964, Prog. Optics, 3, 29.
\bibitem[Kuchner \& Traub, 2002]{kuch02} Kuchner, M.J., \& Traub, W.A. 2002, ApJ, 570, 900.
\bibitem[Labeyrie 1996]{labe96} Labeyrie, A.  1996, {\it A\&AS}, 118, 517.
\bibitem[Leger et al., 1996]{lege96} Leger, A., Mariotti, J-M., Mennesson, B., et al.  1996, Icarus, 123, 249.
\bibitem[Nisenson \& Papaliolios, 2001]{nise01} Nisenson, P., \& Papaliolios, C.  2001 ApJ, 548, L201.
\bibitem[Riaud et al., 2002]{riau02} Riaud, P., Boccaletti, A., Gillet, S., et al.  2002, A\&A, 396, 345. 
\bibitem[Roddier \& Roddier, 1997]{rodd97} Roddier, F., \& Roddier, C.  1997, PASP, 109, 815.
\bibitem[Shealy 2002]{shea02} Shealy, D.L.  2002, {\it SPIE}, 4770, 28S.
\bibitem[Soummer et al., 2003]{soum03} Soummer, R., Aime, C., \& Falloon, P.E  2003, A\&A, 397, 1161.
\bibitem[Spergel, 2001]{sper01} Spergel, D.N.  2001, preprint (astro-ph/0101142).

\end{thebibliography}
\end{document}